\documentclass[runningheads]{llncs}
\usepackage[T1]{fontenc}
\usepackage{graphicx}
\usepackage{amssymb}
\usepackage{bm}
\usepackage{amsmath}
\usepackage[eulergreek]{sansmath}
\usepackage[cal=cm,scr=dutchcal]{mathalfa}

\DeclareFontEncoding{LGR}{}{}
\DeclareSymbolFont{sfgreek}{LGR}{cmss}{m}{n}
\DeclareMathSymbol{\sfbeta}{\mathord}{sfgreek}{`b}
\DeclareMathSymbol{\sfeta}{\mathord}{sfgreek}{`h}
\DeclareMathSymbol{\sfgamma}{\mathord}{sfgreek}{`g}
\DeclareMathSymbol{\sfpi}{\mathord}{sfgreek}{`p}
\DeclareMathSymbol{\sftau}{\mathord}{sfgreek}{`t}
\DeclareMathSymbol{\sfphi}{\mathord}{sfgreek}{`f}
\DeclareMathSymbol{\sfchi}{\mathord}{sfgreek}{`q}
\DeclareMathSymbol{\sfpsi}{\mathord}{sfgreek}{`y}
\DeclareMathSymbol{\sfomega}{\mathord}{sfgreek}{`w}
\DeclareMathSymbol{\sfPhi}{\mathord}{sfgreek}{`F}

\newcommand{\mymathsf}[1]{\mbox{\sansmath$\mathsf{#1}$}}

\begin{document}
%
\title{Towards full `Galilei general relativity': Bargmann-Minkowski and Bargmann-Galilei spacetimes\thanks{This manuscript has been authored by UT-Battelle, LLC, under contract DE-AC05-00OR22725 with the US Department of Energy (DOE). The US government retains and the publisher, by accepting the article for publication, acknowledges that the US government retains a nonexclusive, paid-up, irrevocable, worldwide license to publish or reproduce the published form of this manuscript, or allow others to do so, for US government purposes. DOE will provide public access to these results of federally sponsored research in accordance with the DOE Public Access Plan (\texttt{http://energy.gov/downloads/doe-public-access-plan}).}}
%
%
\author{Christian Y. Cardall\inst{1}\orcidID{0000-0002-0086-105X}}
%
\authorrunning{C. Y. Cardall}
%
\institute{Physics Division, Oak Ridge National Laboratory, Oak Ridge, TN 37831-6354, USA 
\email{cardallcy@ornl.gov}}
\maketitle              
\begin{abstract}
Galilei-Newton spacetime $\mathbb{G}$ with its Galilei group can be understood as a `degeneration' as $c \rightarrow \infty$ of Minkowski spacetime $\mathbb{M}$ with its Poincar\'e group. 
$\mathbb{G}$ does not have a spacetime metric and its Galilei symmetry transformations do not include energy; but Bargmann-Galilei spacetime $B\mathbb{G}$, a 5-dimensional extension that preserves Galilei physics, remedies these infelicities. 
Here an analogous Bargmann-Minkowski spacetime $B\mathbb{M}$ is described.
While not necessary for Poincar\'e physics, it may illuminate a path towards a more extensive `Galilei general relativity' than is presently known, which would be a useful---and conceptually and mathematically sound---approximation in astrophysical scenarios such as core-collapse supernovae.

\end{abstract}
%
%
%
\section{Introduction}
\label{sec:Introduction}

The terms `relativistic physics' and `non-relativistic physics' refer to what might be called something else---perhaps `Poincar\'e relativity' and `Galilei relativity' respectively.
In terms of space, so-called `non-relativistic physics' is in an important sense just as relativistic as `relativistic physics'.
Einstein's essential innovation in so-called `relativistic physics' is not relativity in general, but specifically the relativity of time, or more precisely, the relativity of simultaneity: 
space is mixed into time in Lorentz transformations but not in homogeneous Galilei transformations.
However, time is mixed into space in both Lorentz transformations and homogeneous Galilei transformations.
Thus the presence of `relativity' in both cases---albeit space only in one case, and both space and time in the other---justifies more careful reference to `Galilei relativity' and `Poincar\'e relativity' instead of `non-relativistic physics' and `relativistic physics'.  

Now, what about Einstein circa 1905 vs. Einstein circa 1915?
Having freed the term `relativity' from specific attachment to the world according to Einstein and recognizing its relevance to the world according to Newton and Galilei, the terms `special relativity' and `general relativity' must be reconsidered as well.
The spacetime of Einstein circa 1905 is an affine space, which can be regarded as a flat differentiable manifold.
In contrast, the spacetime of Einstein circa 1915 is a more general pseudo-Riemann manifold whose curvature is determined by the energy and momentum of matter and radiation upon it.

This distinction---between flat and curved spacetime---is what ought to be meant by the terms `special relativity' and `general relativity', without regard for whether the physics is governed by the Poincar\'e group or the Galilei group \cite{Cardall2019Minkowski-and-G}.
In this perspective the key difference is not between `relativistic physics'---whether `special' or `general'---governed by the Poincar\'e group on the one hand, and `non-relativistic physics' governed by the Galilei group on the other.
Instead, what distinguishes `special relativity' from `general relativity' is whether the group in question---whether Poincar\'e, or Galilei---applies to spacetime \textit{globally}, in which case it is an affine space; or only \textit{locally}, in which case its curvature is determined by its energy/momentum/mass content.
The proper references, then, would be to `Poincar\'e special relativity' and `Poincar\'e general relativity', and to `Galilei special relativity' and `Galilei general relativity'.

One might hypothesize that these linguistic shifts, unavoidably associated also with conceptual shifts, point toward a unified perspective on Poincar\'e and Galilei physics that may bear fruit in a Galilei general relativity more extensive than that presently known (described for instance in \cite{de-Saxce2016Galilean-Mechan,de-Saxce20175-Dimensional-T}).
This would play out as follows.
A 5D (5-dimensional) Bargmann extension of 4D Galilei-Newton spacetime is needed to include energy in a tensor formalism on a pseudo-Riemann manifold \cite{de-Saxce2012Bargmann-group-,de-Saxce2016Galilean-Mechan}.
A 5D Bargmann extension of 4D Minkowski spacetime is not \textit{needed}, but it is \textit{allowed}, and may illumniate a path to a \textit{full} Galilei general relativity, in which \textit{full} spacetime curvature (including possible curvature of `space slices') is \textit{determined} by the energy/momentum/mass content (not just given by assumption of Newton's gravitational potential as an ad hoc input).
As a first step towards this goal, a 5D Bargmann extension of 4D Minkowski spacetime that limits nicely to the 5D Bargmann extension of 4D Galilei-Newton spacetime is described here, and the metric tensors of possible corresponding 5D curved spacetime generalizations are displayed.

\section{Minkowski spacetime $\mathbb{M}$}
\label{sec:MinkowskiSpacetime}

	Minkowski spacetime $\mathbb{M}$ is a 4-dimensional affine space with underlying vector space $\mathrm{V}_\mathbb{M}$. 
The invariant structure on $\mathrm{V}_\mathbb{M}$ that governs causality is the null cone, embodied in a 4-metric $\bm{g}$.
With respect to a Minkowski basis $( \bm{e}_0, \bm{e}_1, \bm{e}_2, \bm{e}_3 )$ of $\mathrm{V}_\mathbb{M}$, the metric $\bm{g}$ is represented by
\[
\mathsf{g}
	= \sfeta 
	= \begin{bmatrix} 
		-c^2 & \mymathsf{0} \\
		\mymathsf{0} & \mymathsf{1}
	\end{bmatrix}
	= \begin{bmatrix} 
		-c^2 & 0_j \\
		0_i & 1_{i j}
	\end{bmatrix}.
\]
This `Minkowski matrix' is invariant under Lorentz transformations represented by matrices $\mathsf{P}_\mathbb{M}$: 
\[
\mathsf{P}_\mathbb{M}^\mathrm{T}\, \sfeta\, {\mathsf{P}_\mathbb{M}} = \sfeta.
\]
It is well known that an element $\mathsf{P}_\mathbb{M}^+$ of the identity component of the Lorentz group (restricted Lorentz group) can be factored into a boost and a rotation:
\[
\mathsf{P}_\mathbb{M}^+ = \mathsf{L}_\mathbb{M} \, \mathsf{R}.
\]
Here 
\[
\mathsf{R}
	= \begin{bmatrix}
		1 & \mymathsf{0} \\
		\mymathsf{0} & \mathsf{R}_\mathbb{S}
	\end{bmatrix},
\]
with $\mathsf{R}_\mathbb{S} \in \mathrm{SO}(3)$ a rotation of the subspace $\mathrm{V}_\mathbb{S}$ of $\mathrm{V}_\mathbb{M}$ spanned by $( \bm{e}_1, \bm{e}_2, \bm{e}_3 )$.
A boost is parametrized by a 3-column $\mathsf{u} \in \mathbb{R}^{3 \times 1}$:
\[
\mathsf{L}_\mathbb{M} 
	= \begin{bmatrix}
		\Lambda_\mathsf{u} &&
			 \frac{1}{c^2} \, \Lambda_\mathsf{u} \, \mathsf{u}^\mathrm{T} \\[5pt]
		\Lambda_\mathsf{u} \, \mathsf{u} && \mymathsf{1} + \frac{1}{\lVert \mathsf{u} \rVert^2}( \Lambda_\mathsf{u}  - 1 )\, \mathsf{u} \, \mathsf{u}^\mathrm{T}
	\end{bmatrix},
\]
where
\[
\Lambda_\mathsf{u} = \left( 1 - \frac{\lVert \mathsf{u} \rVert^2}{c^2} \right)^{-1/2}
\]
is the Lorentz factor associated with $\mathsf{u}$, and $\lVert \mathsf{u} \rVert^2 = \mathsf{u}^\mathrm{T} \mathsf{u}$ is the squared Euclid norm with respect to an orthonormal basis of $\mathrm{V}_\mathbb{S}$ (naturally appropriate to a Minkowski basis of $\mathrm{V}_\mathbb{M}$).

The inverse metric $\overleftrightarrow{\bm{g}}$ is represented by
\[
\overleftrightarrow{\sfeta}
	= \begin{bmatrix} 
		-\frac{1}{c^2} & \mymathsf{0} \\[5pt]
		\mymathsf{0} & \mymathsf{1}
	\end{bmatrix}
	= \begin{bmatrix} 
		-\frac{1}{c^2} & \mathsf{0}^j \\[5pt]
		\mathsf{0}^i & \mathsf{1}^{i j}
	\end{bmatrix},
\]
and is also invariant, according to
\[
\mathsf{P}_\mathbb{M}^{-1} \, \overleftrightarrow{\sfeta}\, \mathsf{P}_\mathbb{M}^{-\mathrm{T}} = \overleftrightarrow{\sfeta}.
\]
Given $\bm{g}$ there is metric duality between vectors and linear forms (`raising and lowering of indices').
Let ${\mathrm{V}_\mathbb{M}}_*$ be the vector space of linear forms on $ \mathrm{V}_\mathbb{M}$.
For $\bm{a} \in \mathrm{V}_\mathbb{M}$ and $\bm{\omega} \in {\mathrm{V}_\mathbb{M}}_*$,
\begin{eqnarray}
\underline{\bm{a}} 
	&=& \bm{g}(\bm{a},\cdot) \in {\mathrm{V}_\mathbb{M}}_* \nonumber \\
\overleftarrow{\bm{\omega}} 
	&=& \overleftrightarrow{\bm{g}}(\bm{\omega},\cdot) \in \mathrm{V}_\mathbb{M}. \nonumber
\end{eqnarray}
For natural contractions (here, never scalar products!), use the dot operator, for example 
\[
\begin{array}{rllll}
	\underline{\bm{a}} & = & \bm{g} \cdot \bm{a} & = &  \bm{a} \cdot \bm{g}, \\[5pt]
	\overleftarrow{\bm{\omega}} & = & \bm{\omega} \cdot \overleftrightarrow{\bm{g}} & = & \overleftrightarrow{\bm{g}} \cdot \bm{\omega}, \\[5pt]
	\overleftarrow{\bm{F}} & = & \overleftrightarrow{\bm{g}} \cdot \bm{F}, & & \\
	\overleftrightarrow{\bm{F}} & = & \overleftrightarrow{\bm{g}} \cdot \bm{F} \cdot \overleftrightarrow{\bm{g}} & = & \overleftarrow{\bm{F}} \cdot \overleftrightarrow{\bm{g}}.
\end{array}
\]

\section{Galilei-Newton spacetime $\mathbb{G}$}
\label{sec:GalileiNewtonSpacetime}

	Galilei-Newton spacetime $\mathbb{G}$ is a `degeneration' of Minkowski spacetime $\mathbb{M}$ as $c \rightarrow \infty$.
The metric $\bm{g}$ asymptotes (without a true limit) in a manner that suggests that the linear form $\bm{\tau} = \bm{\mathrm{d}}t$, where $t$ is the time coordinate, becomes the invariant structure governing causality, embodying absolute time.
With respect to what will be called a Galilei basis, the `time form' $\bm{\tau}$ is represented by
\[
\sftau = \begin{bmatrix} 1 & \mymathsf{0} \end{bmatrix} = \begin{bmatrix} 1 & 0_i \end{bmatrix}.
\]
The inverse metric $\overleftrightarrow{\bm{g}}$ limits sensibly to another invariant structure, the $(2,0)$ tensor $\overleftrightarrow{\bm{\gamma}}$.
With respect to a Galilei basis, $\overleftrightarrow{\bm{\gamma}}$ is represented by
\[
\overleftrightarrow{\sfgamma}
	= \begin{bmatrix} 
		0 & \mymathsf{0} \\
		\mymathsf{0} & \mymathsf{1}
	\end{bmatrix}
	= \begin{bmatrix} 
		0 & 0^j \\
		0^i & 1^{i j}
	\end{bmatrix}.
\]
	The covector $\bm{\tau}$ is invariant according to
\[	
\begin{bmatrix} 1 & 0_i \end{bmatrix} \mathsf{P}_\mathbb{G} 
	=  \begin{bmatrix} 1 & 0_i \end{bmatrix},
\]	
and the tensor $\overleftrightarrow{\bm{\gamma}}$ is invariant according to
\[
\mathsf{P}_\mathbb{G}^{-1} \, \begin{bmatrix} 
		0 & \mathsf{0}^j \\
		\mathsf{0}^i & \mathsf{1}^{i j}
	\end{bmatrix} \, \mathsf{P}_\mathbb{G}^{-\mathrm{T}}
	= \begin{bmatrix} 
		0 & \mathsf{0}^j \\
		\mathsf{0}^i & \mathsf{1}^{i j}
	\end{bmatrix}.
\]
Here the homogeneous Galilei transformations $\mathsf{P}_\mathbb{G}$ are the $c \rightarrow \infty$ limit of the Lorentz transformations $\mathsf{P}_\mathbb{M}$.
As with the restricted Lorentz group, elements $\mathsf{P}_\mathbb{G}^+$ of the identity component of the homogeneous Galilei group can be factored into a boost and a rotation: 
\[
\mathsf{P}_\mathbb{G}^+ = \mathsf{L}_\mathbb{G} \, \mathsf{R}.
\]
Here $\mathsf{R}$ is the same as before, and the Galilei boost is
\[
\mathsf{L}_\mathbb{G} 
	= \begin{bmatrix}
		1 & \mymathsf{0} \\
		\mathsf{u} & \mymathsf{1}
	\end{bmatrix}.
\]
The tensor $\overleftrightarrow{\bm{\gamma}}$ derived from $\overleftrightarrow{\bm{g}}$ does not qualify as an inverse metric tensor on $\mathrm{V}_\mathbb{G}$.
It has no inverse because it is degenerate:
\[
\overleftrightarrow{\bm{\gamma}} ( \bm{\tau}, \cdot ) 
	= \bm{\tau} \cdot \overleftrightarrow{\bm{\gamma}} = 0.
\]
There is no spacetime metric on $\mathbb{G}$.
Tensor algebra is more constrained: there is no metric duality---no `raising and lowering of indices'.
There are spacetime tensors but they can only be of fixed type.

\section{Decomposition of $\mathbb{M}$ and $\mathbb{G}$ into time and space}
\label{sec:Decomposition_M_G}

Theories formulated in terms of tensors on spacetime can only be compared with experiments once spacetime is broken into `time' and `space' (and tensors are decomposed accordingly).

	Affine spacetimes permit `inertial observers' with straight worldlines and no rotation.
The splitting of space and time as perceived by a single inertial observer is formally similar on $\mathbb{M}$ and $\mathbb{G}$.
Select an event $\mathbf{O}$ of $\mathbb{M}$ or $\mathbb{G}$ as origin.
Select a Minkowski basis of $\mathrm{V}_\mathbb{M}$ or a Galilei basis of $\mathrm{V}_\mathbb{G}$, designated $\left( \bm{e}_\mu \right) = \left( \bm{e}_0, \bm{e}_1, \bm{e}_2, \bm{e}_3 \right)$.
A point $\mathbf{X} \in \mathbb{M}, \mathbb{G}$ is given in terms of coordinates $( X^\mu) = ( t, x^i )$ by
\[
\mathbf{X} = \mathbf{O} + \bm{e}_\alpha \, X^\alpha.
\]
The time axis $\mathbb{T}$ is the straight line
\[
\mathbb{T} = \{\mathbf{O} +  \bm{e}_0 \, t \mid t \in \mathbb{R} \}.
\]
Interpret $\mathbb{T}$ as as the worldline of a fiducial (and inertial) observer whose tangent vector is the constant 4-velocity $\bm{n} = \bm{e}_0$.
Let $\mathrm{V}_\mathbb{S}$ be the subspace of $\mathrm{V}_\mathbb{M}$ or $\mathrm{V}_\mathbb{G}$ spanned by $\left( \bm{e}_1, \bm{e}_2, \bm{e}_3 \right)$.
For a given time $t \in \mathbb{R}$, consider a one-to-one mapping 
\begin{eqnarray}
V_\mathbb{S} & \rightarrow & \mathbb{M} \mathrm{\ or \ } \mathbb{G} \nonumber \\
\bm{x} & \mapsto & \mathbf{O} +  \bm{n} \, t + \bm{x}. \nonumber
\end{eqnarray} 
The image of this mapping is a hyperplane $\mathbb{S}_t$ through the event $\mathbf{O} +  \bm{n} \, t $: 
\[
\mathbb{S}_t = \{ \mathbf{O} + \bm{n} \, t + \bm{e}_i \, x^i \mid ( x^i ) \in \mathbb{R}^3 \}.
\]
$\mathbb{S}_t$ is a 3-dimensional affine subspace of $\mathbb{M}$ or $\mathbb{G}$ with underlying vector space $\mathbb{V}_\mathbb{S}$.
Interpret $\mathbb{S}_t$ as `space' according to the fiducial observer at her time $t$---a surface of `simultaneity'.
It is evident from the factorization $\mathsf{P}^+ = \mathsf{L} \, \mathsf{R}$ that $\mathbb{V}_\mathbb{S}$ is rotationally invariant. Thus $\mathrm{V}_\mathbb{S}$ is endowed with a Euclid metric $\bm{\gamma}$ defining the usual scalar product on $\mathbb{R}^3$.
Each hypersurface $\mathbb{S}_t$ is a level surface of the coordinate function $t$.
The complete collection $\left( \mathbb{S}_t \right)_{t \in \mathbb{R}}$ is said to be a foliation of $\mathbb{M}$ or $\mathbb{G}$.

\section{A material particle on $\mathbb{M}$ and $\mathbb{G}$}
\label{sec:MaterialParticle_M_G}

	A material particle is represented by a timelike curve $\mathbf{X}(\tau)$ in spacetime, parametrized by the particle's proper time $\tau$.
The tangent vector $\bm{U}(\tau) = \mathrm{d} \mathbf{X} / \mathrm{d} \tau$, the 4-velocity, satisfies $\bm{g}(\bm{U},\bm{U}) = -c^2$ on $\mathbb{M}$ and $\bm{\tau}(\bm{U}) = 1$ on $\mathbb{G}$.
Select a fiducial observer with global coordinates $( t, x^i )$ associated with a choice of origin $\mathbf{O}$ of $\mathbb{M}$ or $\mathbb{G}$ and a Minkowski or Galilei basis $( \bm{n}, \bm{e}_i )$ for $\mathrm{V}_\mathbb{M}$ or $\mathrm{V}_\mathbb{G}$.
Decompose $\bm{U}$ into measurable pieces parallel to $\mathbb{T}$ and tangent to $\mathbb{S}_t$:
\[
\bm{U} = \frac{\mathrm{d}t}{\mathrm{d}\tau} \, \frac{\mathrm{d}\mathbf{X}}{\mathrm{d}t}
	 	= \frac{\mathrm{d}t}{\mathrm{d}\tau} \left( \bm{n} + \bm{v} \right).
\]
This follows from the 4-column representations
\[
\mathsf{X} = \begin{bmatrix} t \\ \mathsf{x} ( t )  \end{bmatrix},
 \ \ \ \mathsf{n} = \begin{bmatrix} 1 \\ \mymathsf{0} \end{bmatrix},
 \ \ \ \mathsf{v} = \begin{bmatrix} 0 \\ \mathrm{d} \mathsf{x} / \mathrm{d}t \end{bmatrix}.
\]
The leading factor $\mathrm{d}t / \mathrm{d}\tau$ is determined by the fundamental structures $\bm{g}$ and $\bm{\tau}$ governing causality.
Proper time increments $\mathrm{d} \tau$ are given by
\[
\begin{array}{rclclcl}
c \, \mathrm{d}\tau 
	&=& \sqrt{ - \bm{g} \left( \mathrm{d}\mathbf{X}, \mathrm{d}\mathbf{X} \right)}
	&=& c \, \Lambda_{\bm{v}}^{-1} \, \mathrm{d}t 
	& &  (\mathrm{on \ } \mathbb{M}),  \\[5pt]
\mathrm{d}\tau &=& \bm{\tau} \left( \mathrm{d}\mathbf{X} \right) 
	&=& \mathrm{d} t 
	& &  (\mathrm{on \ } \mathbb{G}),
\end{array}
\]
so that $\mathrm{d}t / \mathrm{d}\tau = \Lambda_{\bm{v}}$ on $\mathbb{M}$ and $\mathrm{d}t / \mathrm{d}\tau = 1$ on $\mathbb{G}$.

So far so good on both $\mathbb{M}$ and $\mathbb{G}$: a spacetime description of particle kinematics---specifying where a particle is (a point $\mathbf{X}(\tau)$ on its worldline), and how fast it is moving (the 4-velocity $\bm{U}$ tangent to the worldline)---is unproblematic in either case.

However, a spacetime formulation of particle dynamics turns out to be more problematic on $\mathbb{G}$.
Because of the absence of a spacetime metric there is no equivalence between inertia and total energy.
The best one can do is include kinetic energy in the time component of a `relative energy momentum covector' $\bm{\Pi}$ \cite{Cardall2020Combining-3-Mom}. 
But this is not fully satisfying because the notion of kinetic energy (energy of motion) inherently depends on a choice of observer (motion relative to whom?): 
the fiducial observer covector $\underline{\bm{n}}$ is built into the definition of the 4-covector $\bm{\Pi}$ whose time component is the kinetic energy relative to the fiducial observer.
The unsatisfying result is that Lorentz or homogeneous Galilei transformations cannot transform the components of $\bm{\Pi}$ in such a way as to demonstrate the transformation rule of kinetic energy.
This motivates extensions of the Lorentz and homogeneous Galilei groups that address the transformation of kinetic energy.

\section{Bargmann spacetimes $B\mathbb{M}$ and $B\mathbb{G}$}
\label{sec:BargmannSpacetimes}

Work backwards towards Bargmann-Minkowski (or B-Minkowski) spacetime $B \mathbb{M}$ and Bargmann-Galilei (or B-Galilei) spacetime $B \mathbb{G}$ by considering a `5-velocity'  $\hat{\bm{U}}$ that extends the 4-velocity $\bm{U}$ on $\mathbb{M}$ or $\mathbb{G}$.
The fifth component will be the specific kinetic energy---kinetic energy per unit mass---involving only 3-velocity.
With respect to a B-Minkowski or B-Galilei basis (fiducial observer):
\[
\hat{\mathsf{U}}
		= \begin{bmatrix} \Lambda_\mathsf{v}  \, \phantom{\mathsf{v}} 
				\\ \Lambda_\mathsf{v} \, \mathsf{v}
				\\ c^2 \left( \Lambda_\mathsf{v} - 1 \right)
			\end{bmatrix} 
			\quad (\mathrm{on \ } B\mathbb{M}), \quad \quad
\hat{\mathsf{U}}
		= \begin{bmatrix} 1
				\\ \mathsf{v}
				\\ \frac{1}{2} \lVert \mathsf{v} \rVert^2
			\end{bmatrix} 
			\quad (\mathrm{on \ } B\mathbb{G}).
\]
The additional dimension requires an additional coordinate. 
A point $\hat{\mathbf{X}} (\tau)$ along the particle worldline is represented by a 5-column
\[
\hat{\mathsf{X}} = \begin{bmatrix} t \\ \mathsf{x} ( t ) \\ \eta ( t ) \end{bmatrix} 
	= \begin{bmatrix} t \\ x^i ( t ) \\ \eta ( t ) \end{bmatrix}.
\]
The proper time $\tau$ is governed by $\bm{g}$ or $\bm{\tau}$ as before;
these are now regarded as tensors on $B\mathbb{M}$ or $B\mathbb{G}$ respectively.
The fifth component $\hat{U}^\eta$ of the 5-velocity $\hat{\bm{U}} = \mathrm{d} \hat{\mathbf{X}} / \mathrm{d}\tau$ must satisfy
\begin{eqnarray}
\hat{U}^\eta
	= \frac{ \mathrm{d} \eta }{ \mathrm{d} \tau } 
	=  \frac{ \mathrm{d} t }{ \mathrm{d} \tau } \frac{ \mathrm{d} \eta }{ \mathrm{d} t }
	&=& c^2 \left( \Lambda_\mathsf{v} - 1 \right) \ \ \ (\mathrm{on \ } B\mathbb{M}), 
			\nonumber \\
	&=& \frac{1}{2} \lVert \mathsf{v} \rVert^2 \ \ \ \ \ \ \ \ \ (\mathrm{on \ } B\mathbb{G}). 
\label{eq:ActionCoordinateRelation}
\end{eqnarray}
It is apparent that $\eta$ has units of action/mass; call it the `action coordinate'.
The above `action coordinate relation' will prove crucial to the geometry of $B\mathbb{M}$ and $B\mathbb{G}$.

Next, determine the $5 \times 5$ B-Lorentz transformation matrices $\hat{\mathsf{P}}_{B\mathbb{M}}^+$ and homogeneous B-Galilei transformation matrices $\hat{\mathsf{P}}_{B\mathbb{G}}^+$ that appear in the 5-velocity transformation
\[
\hat{\mathsf{U}} 
	= \hat{\mathsf{P}}^+ \, \hat{\mathsf{U}}',
\]
or in (4+1)-dimensional form
\[
\begin{bmatrix} \mathsf{U} \\ U^\eta \end{bmatrix} 
	= \begin{bmatrix} \mathsf{P}^+ & \mymathsf{0} \\ \sfPhi & 1 \end{bmatrix}
		\begin{bmatrix} \mathsf{U}' \\ U'^\eta \end{bmatrix}. 
\]
The 4-column $\mymathsf{0} = \begin{bmatrix} 0^\mu \end{bmatrix}$ in
$\hat{\mathsf{P}}^+$ ensures that the 4D relation $\mathsf{U} = \mathsf{P}^+ \, \mathsf{U}'$ on $\mathbb{M}$ or $\mathbb{G}$ is preserved when embedded in the 5D setting of $B\mathbb{M}$ or $B\mathbb{G}$.
It also ensures that the matrix representations of $\bm{g}$ and $\bm{\tau}$ do not acquire non-vanishing components in the $\eta$ dimension when these are regarded as tensors on  
$B\mathbb{M}$ and $B\mathbb{G}$.
This means that the `timelike 4-velocity' character of $\bm{U}$ on $\mathbb{M}$ or $\mathbb{G}$ is preserved when it is extended to the 5-velocity $\hat{\bm{U}}$ on $B\mathbb{M}$ or $B\mathbb{G}$.
The 4-row $\sfPhi$ in
$\hat{\mathsf{P}}^+$ is determined by the requirement that the fifth component of the above transformation of $\hat{\mathsf{U}}$ yield the transformation rule for (specific) kinetic energy.
For Poincar\'e physics this can be derived most easily from the time component of the 4D relation $\mathsf{U} = \mathsf{P}_\mathbb{M}^+ \, \mathsf{U}'$.
For Galilei physics it is derived from the transformed 3-velocity (Galilei velocity addition with rotation), the space components of the 4D relation $\mathsf{U} = \mathsf{P}_\mathbb{G}^+ \, \mathsf{U}'$.
The resulting expressions for $\sfPhi$ are
\begin{eqnarray}
\sfPhi &=& \begin{bmatrix} c^2 \left( \Lambda_\mathsf{u} - 1 \right) 
			&& \Lambda_\mathsf{u} \, \mathsf{u}^\mathrm{T} \, \mathsf{R}_\mathbb{S}\end{bmatrix} 
		\quad (\mathrm{on \ } B\mathbb{M}), \quad \quad \nonumber \\
 	&=& \begin{bmatrix}  \phantom{c^2 \,} \frac{1}{2} \lVert \mathsf{u} \rVert^2 \phantom{ - 1}
			&& \phantom{\Lambda_\mathsf{u} }
				\mathsf{u}^\mathrm{T} \, \mathsf{R}_\mathbb{S}\end{bmatrix} 
		\quad (\mathrm{on \ } B\mathbb{G}).
\label{eq:Cocycle}
\end{eqnarray}
No new parameters beyond $\mathsf{u} \in \mathbb{R}^{3 \times 1}$ and $\mathsf{R}_\mathbb{S} \in \mathrm{SO}(3)$ already present in a Lorentz transformation $\mathsf{P}_{\mathbb{M}}^+$ or homogeneous Galilei transformation $\mathsf{P}_{\mathbb{G}}^+$ are introduced.

The set of B-Lorentz transformations $\hat{\mathsf{P}}_{B\mathbb{M}}^+$ and the set of homogeneous B-Galilei transformations $\hat{\mathsf{P}}_{B\mathbb{G}}^+$ are subgroups of $\mathrm{GL}(5)$.
It is evident that these sets of matrices contain the identity ($\mathsf{u} = \mymathsf{0}$ and $\mathsf{R}_\mathbb{S} = \mymathsf{1}$).
Once again there is a factorization $\hat{\mathsf{P}}^+ = \hat{\mathsf{L}} \, \hat{\mathsf{R}}$, so that inverses are given by ${{}\hat{\mathsf{P}}^+}^{-1} = \hat{\mathsf{R}}^\mathrm{T} \, \hat{\mathsf{L}}^{-1}$ with $\hat{\mathsf{L}}^{-1}$ obtained from $\hat{\mathsf{L}}$ via $\mathsf{u} \mapsto -\mathsf{u}$.
Closure under matrix multiplication is shown by considering the product
\[
{{}\hat{\mathsf{P}}^+}'' = {{}\hat{\mathsf{P}}^+} \, {{}\hat{\mathsf{P}}^+}',
\]
or
\[
\begin{bmatrix} {\mathsf{P}^+}'' & \mymathsf{0} \\ \sfPhi'' & 1 \end{bmatrix}
	= \begin{bmatrix} {\mathsf{P}^+} & \mymathsf{0} \\ \sfPhi & 1 \end{bmatrix}
		\begin{bmatrix} {\mathsf{P}^+}' & \mymathsf{0} \\ \sfPhi' & 1 \end{bmatrix} 
	= \begin{bmatrix} {\mathsf{P}^+} \, {\mathsf{P}^+}' && \mymathsf{0} 
		\\ \sfPhi \, {\mathsf{P}^+}' + \sfPhi' && 1 \end{bmatrix}. 
\]
The $4 \times 4$ matrix relation 
\begin{equation}
{\mathsf{P}^+}'' = {\mathsf{P}^+} \, {\mathsf{P}^+}'
\label{eq:LorentzGalileiClosure}
\end{equation}
in the upper-left block is simply the known closure of the restricted Lorentz or homogeneous Galilei group.
The remaining question is whether the 4-row
\[
\sfPhi'' = \sfPhi \, {\mathsf{P}^+}' + \sfPhi' 
\]
is in the form of Eq.~(\ref{eq:Cocycle}), with the relevant expressions involving $\mathsf{u}''$ and $\mathsf{R}''$ determined consistently from Eq.~(\ref{eq:LorentzGalileiClosure}).
Direct computation shows that the answer is yes, completing the demonstration of closure.

The existence of a `Bargmann metric' $\bm{G}$ is suggested by the `action coordinate relation' in Eq.~(\ref{eq:ActionCoordinateRelation}) relating coordinate variations along a material particle worldline, and it turns out to be invariant under B-Lorentz or homogeneous B-Galilei transformations, making it a fundamental structure on $B\mathbb{M}$ or $B\mathbb{G}$.
On $B\mathbb{M}$, use $\Lambda_\mathsf{v} = \mathrm{d} t / \mathrm{d} \tau$ and $c^2 \, \mathrm{d} \tau^2 = c^2 \, \mathrm{d} t^2 - \lVert \mathrm{d} \mathsf{x} \rVert^2$ in Eq.~(\ref{eq:ActionCoordinateRelation}) to deduce
\[
- 2 \, \mathrm{d} \eta \, \mathrm{d} t + \mathrm{d} x^a \, 1_{a b} \, \mathrm{d} x^b 
	+ \frac{1}{c^2} \, \mathrm{d} \eta^2 = 0 \ \ \ (\mathrm{on \ } B\mathbb{M}).
\]
On $B\mathbb{G}$, use $\mathrm{d} \tau = \mathrm{d} t$ and $\lVert \mathsf{v} \rVert^2 \, \mathrm{dt}^2 =  \lVert \mathrm{d} \mathsf{x} \rVert^2$ to deduce analogously
\[
- 2 \, \mathrm{d} \eta \, \mathrm{d} t + \mathrm{d} x^a \, 1_{a b} \, \mathrm{d} x^b 
	 = 0 \ \ \ (\mathrm{on \ } B\mathbb{G}).
\]
In both cases the left-hand side looks like a line element, suggestive of a Bargmann metric (or B-metric) $\bm{G}$ represented by the B-Minkowski or B-Galilei matrix
\[
\mathsf{G}
	= \hat{\sfeta}_{B\mathbb{M}}
	= \begin{bmatrix} 0 & 0_j & -1 \\[5pt]
		 0_i & 1_{ij} & 0_i \\[5pt]
		  -1 & 0_j & \frac{1}{c^2} \end{bmatrix} 
		 \ \ \ (\mathrm{on \ } B\mathbb{M}), \quad \quad
\mathsf{G}
	= \hat{\sfeta}_{B\mathbb{G}}	
	= \begin{bmatrix} 0 & 0_j & -1 \\[5pt] 0_i & 1_{ij} & 0_i \\[5pt] -1 & 0_j & 0 \end{bmatrix} 
		=  \ \ \ (\mathrm{on \ } B\mathbb{G})
\]
with respect to a B-Minkowski or B-Galilei basis.
The invariance condition reads
\[
\hat{\mathsf{P}}^\mathrm{T} \, \hat{\sfeta} \, \hat{\mathsf{P}}
	=  \hat{\sfeta}
\]
and is verified by direct computation for both $\hat{\mathsf{P}}_{B\mathbb{M}}^+$ with $\hat{\sfeta}_{B\mathbb{M}}$ and $\hat{\mathsf{P}}_{B\mathbb{G}}^+$ with $\hat{\sfeta}_{B\mathbb{G}}$.
However, the 6-dimensional Lie groups of B-Lorentz and homogeneous B-Minkowski transformations are only subgroups of the 10-dimensional Lie groups that preserve $\bm{G}$ for $B\mathbb{M}$ and $B\mathbb{G}$ respectively.

The above calculation suggesting the existence of $\bm{G}$ also shows that 
\[
\bm{G} \left( \hat{\bm{U}}, \hat{\bm{U}} \right) 
	= 	\hat{\mathsf{U}}^\mathrm{T} \, \hat{\sfeta} \, \hat{\mathsf{U}}
	=	0,
\]
that is, that $\hat{\bm{U}}$ is null with respect to $\bm{G}$.
This is so even though $\hat{\bm{U}}$ remains timelike with respect to $\bm{g}$ or $\bm{\tau}$ as appropriate, as noted previously.

The inverse metric $\overleftrightarrow{\bm{G}}$ is represented by
\[
\overleftrightarrow{\mathsf{G}}
	= \overleftrightarrow{\hat{\sfeta}}_{B\mathbb{M}} 
	= \begin{bmatrix}  -\frac{1}{c^2} & 0^j & -1 \\[5pt] 
					0^i & 1^{ij} & 0^i \\[5pt]
		 			-1 & 0^j & 0 \end{bmatrix} 
		 \ \ \ (\mathrm{on \ } B\mathbb{M}), \quad
\overleftrightarrow{\mathsf{G}}
	= \overleftrightarrow{\hat{\sfeta}}_{B\mathbb{G}} 
	= \begin{bmatrix} 0 & 0^j & -1 \\[5pt]
					 0^i & 1^{ij} & 0^i \\[5pt]
					  -1 & 0^j & 0 \end{bmatrix} 
		 \ \ \ (\mathrm{on \ } B\mathbb{G}) \nonumber
\]
with respect to a B-Minkowski or B-Galilei basis.

Note the remarkable difference in the relationship between $\mathbb{M}$ and $\mathbb{G}$ on the one hand and between $B\mathbb{M}$ and $B\mathbb{G}$ on the other, including startlingly different geometric consequences.
Whereas the spacetime $\mathbb{M}$ is a pseudo-Riemann manifold with metric $\bm{g}$ and inverse $\overleftrightarrow{\bm{g}}$, the spacetime $\mathbb{G}$ obtained as $c \rightarrow \infty$ is not:  instead of a metric and its true inverse, one is left with an invariant time form $\bm{\tau}$ and an invariant degenerate inverse `metric' $\overleftrightarrow{\bm{\gamma}}$.
In contrast both $B\mathbb{M}$ and $B\mathbb{G}$ are pseudo-Riemann manifolds with a (flat) metric $\bm{G}$ and inverse $\overleftrightarrow{\bm{G}}$, the versions of both of these on $B\mathbb{M}$ limiting smoothly to those on $B\mathbb{G}$ as $c \rightarrow \infty$, as is evident from the above expressions relative to B-Minkowski and B-Galilei bases.

\section{Conclusion}
\label{sec:Conclusion}

This account of Bargmann-Minkowski spacetime $B\mathbb{M}$ with its metric $\bm{G}$, deduced from Eq.~(\ref{eq:ActionCoordinateRelation}), extends to Poincar\'e physics an ingenious elementary introduction given by de Saxc\'e and Vall\'ee \cite{de-Saxce2016Galilean-Mechan} of the Bargmann group as an extension to the Galilei group.
This approach to the Bargmann group is simple and direct in comparison with its origins in the study of projective representations of Lie groups in quantum mechanics \cite{Bargmann1954On-Unitary-ray-}, but the necessity of transforming kinetic energy is a shared underlying motivation: the Hamiltonian in the `non-relativistic' (forgive the lapse) Schr\"odinger equation contains kinetic energy, and this equation cannot be shown to be Galilei covariant without taking the projective phase into account \cite{Levy-Leblond1976Quantum-fact-an}.

The next step is to relax the assumption of an affine space, allowing instead spacetime curvature determined by its energy-momentum content.
Call the 4D spacetime of standard general relativity `Einstein spacetime' $\mathcal{E}$; in its $3+1$ formulation \cite{Gourgoulhon201231-Formalism-in} in terms of the lapse function $\alpha$, shift 3-vector $\bm{\beta}$, and 3-metric $\bm{\gamma}$, the Lorentz factor of a material particle is $\Lambda = \alpha \, \mathrm{d}t / \mathrm{d}\tau$ and proper time intervals are given by $c^2 \, \mathrm{d}\tau^2 = c^2 \alpha^2 \, \mathrm{d}t^2 - \bm{\gamma} \left( \mathrm{d} \bm{x} + \bm{\beta} \, \mathrm{d}t, \mathrm{d} \bm{x} + \bm{\beta} \, \mathrm{d}t \right)$. 
Using these expressions in Eq.~(\ref{eq:ActionCoordinateRelation}) yields
\[
\beta_a \beta^a \, \mathrm{d}t^2 - 2 \, \mathrm{d}t \, \beta_a \mathrm{d}x^a - 2 \, \alpha \, \mathrm{d} \eta \, \mathrm{d} t + \mathrm{d} x^a \, \gamma_{a b} \, \mathrm{d} x^b 
	+ \frac{1}{c^2} \, \mathrm{d} \eta^2 = 0 \ \ \ (\mathrm{on \ } B\mathcal{E}),
\]
suggestive of a 5D Bargmann-Einstein spacetime $B\mathcal{E}$ with metric $\bm{G}$ and inverse $\overleftrightarrow{\bm{G}}$ represented by
\[
\mathsf{G}
	= \begin{bmatrix} \beta_a \beta^a && \beta_j && -\alpha \\[5pt]
		 \beta_i && \gamma_{ij} && 0_i \\[5pt]
		  -\alpha && 0_j && \frac{1}{c^2} \end{bmatrix}, \quad \quad
\overleftrightarrow{\mathsf{G}}
	= \begin{bmatrix} -\frac{1}{c^2  \alpha^2} && \frac{1}{c^2  \alpha^2} \, \beta^j && -\frac{1}{\alpha} \\[5pt]
					  \frac{1}{c^2  \alpha^2} \, \beta^i && \gamma^{ij} -  \frac{1}{c^2  \alpha^2} \, \beta^i \beta^j&& \frac{1}{\alpha} \, \beta^i \\[5pt]
					  -\frac{1}{\alpha} && \frac{1}{\alpha} \, \beta^j  && 0 \end{bmatrix} 
		 \ \ \ (\mathrm{on \ } B\mathcal{E}). \nonumber
\]
As $c \rightarrow \infty$ this limits smoothly to
\[
\mathsf{G}
	= \begin{bmatrix} \beta_a \beta^a && \beta_j && -\alpha \\[5pt]
		 \beta_i && \gamma_{ij} && 0_i \\[5pt]
		  -\alpha && 0_j && 0 \end{bmatrix}, \quad \quad
\overleftrightarrow{\mathsf{G}}
	= \begin{bmatrix} 0 && 0^j && -\frac{1}{\alpha} \\[5pt]
					  0^i && \gamma^{ij} && \frac{1}{\alpha} \, \beta^i \\[5pt]
					  -\frac{1}{\alpha} && \frac{1}{\alpha} \, \beta^j  && 0 \end{bmatrix} 
		 \ \ \ (\mathrm{on \ } B\mathcal{G}), \nonumber
\]
suggestive of a hitherto unknown `Galilei general relativistic' spacetime $B\mathcal{G}$.
(In both cases these reduce to the previous expressions on $B\mathbb{M}$ and $B\mathbb{G}$ as $\alpha \rightarrow 1$ and $\bm{\beta} \rightarrow 0$.)
Thus there is a reasonable prospect that recasting the $3+1$ formulation of the Einstein equations on $\mathcal{E}$ as a $1+3+1$ formulation on $B\mathcal{E}$ and taking the $c \rightarrow \infty$ limit could yield a Galilei gravitation of enhanced strength in which energy density and stress contribute as sources and give rise to space as well as spacetime curvature, beyond the flat space slices and spacetime curvature determined by mass density alone in Cartan's reformulation of Newtonian gravitation.
This would be a useful---and conceptually and mathematically sound---approximation in astrophysical scenarios such as core-collapse supernovae, in which the energy density and pressure of the nascent neutron star contribute to enhanced gravity at the 10-20\% level, but for which the computationally/numerically fraught phenomena of `Minkowski' bulk fluid flow and back-reaction of gravitational radiation are much less significant.


\subsubsection{Acknowledgements}
Thanks to G\'ery de Saxc\'e for pointing out that preservation of the B-metric $\bm{G}$ is not sufficient to prove closure of the B-Lorentz transformations $\hat{\mathsf{P}}_{B\mathbb{M}}^+$, but that closure directly follows instead from relations obtained from closure of the Lorentz group.

%
%
%

\begin{thebibliography}{1}
\providecommand{\url}[1]{\texttt{#1}}
\providecommand{\urlprefix}{URL }
\providecommand{\doi}[1]{https://doi.org/#1}

\bibitem{Bargmann1954On-Unitary-ray-}
Bargmann, V.: {On Unitary Ray Representations of Continuous Groups}. Annals
  Math.  \textbf{59},  1--46 (1954)

\bibitem{Cardall2019Minkowski-and-G}
{Cardall}, C.Y.: {Minkowski and Galilei/Newton Fluid Dynamics: A Geometric 3+1
  Spacetime Perspective}. Fluids  \textbf{4}, ~1 (2019)

\bibitem{Cardall2020Combining-3-Mom}
{Cardall}, C.Y.: {Combining 3-Momentum and Kinetic Energy on Galilei/Newton
  Spacetime}. Symmetry  \textbf{12}(11), ~1775 (2020)

\bibitem{de-Saxce20175-Dimensional-T}
{de Saxc\'e}, G.: {5-Dimensional Thermodynamics of Dissipative Continua}. In:
  {Fr\'emond}, M., {Maceri}, F., {Vairo}, G. (eds.) Models, Simulation, and
  Experimental Issues in Structural Mechanics, Springer Series in Solid and
  Structural Mechanics, vol.~8, pp. 1--40. Springer, Cham (2017)

\bibitem{de-Saxce2012Bargmann-group-}
{de Saxc\'e}, G., {Vall\'ee}, C.: {Bargmann group, momentum tensor and Galilean
  invariance of Clausius-Duhem inequality}. International Journal of
  Engineering Science  \textbf{50},  216--232 (2012)

\bibitem{de-Saxce2016Galilean-Mechan}
{de Saxc\'e}, G., {Vall\'ee}, C.: Galilean Mechanics and Thermodynamics of
  Continua. John Wiley \& Sons, Inc., Hoboken (2016)

\bibitem{Gourgoulhon201231-Formalism-in}
{Gourgoulhon}, E.: {3+1 Formalism in General Relativity: Bases of Numerical
  Relativity}, Lecture Notes in Physics, vol.~846. Springer, Berlin Heidelberg
  (2012)

\bibitem{Levy-Leblond1976Quantum-fact-an}
{L{\'e}vy-Leblond}, J.M.: {Quantum fact and classical fiction: Clarifying
  Land{\'e}'s pseudo‐paradox}. Am. J. Phys.  \textbf{44},  1130--1132 (1976)

\end{thebibliography}
%

%
%
%
%
%

\end{document}